\documentclass[aps,prb,twocolumn,superscriptaddress]{revtex4}
\usepackage{graphicx}
\usepackage{amssymb}
\usepackage{subfigure}
\usepackage{color}
\usepackage{multirow}
\usepackage[paper=letterpaper,left=1in,right=1in,top=1in,bottom=1in]{geometry}

\begin{document}

\title{A new approach for efficient simulation of Coulomb interactions in ionic fluids}
\author{Natalia A. Denesyuk} 
\email{denesyuk@umd.edu}
\affiliation{Institute for Physical Science and Technology, University of Maryland, College Park, 
Maryland 20742}
\author{John D. Weeks}
\email{jdw@ipst.umd.edu}
\affiliation{Institute for Physical Science and Technology, University of Maryland, College Park, 
Maryland 20742}
\affiliation{Department of Chemistry and Biochemistry, University of Maryland, College Park, 
Maryland 20742}
\date{\today}

\begin{abstract}
We propose a simplified version of local molecular field (LMF) theory to treat Coulomb 
interactions in simulations of ionic fluids. LMF theory relies on splitting the Coulomb potential 
into a short-ranged part that combines with other short-ranged core interactions and
is simulated explicitly. The averaged effects of the remaining long-ranged part are 
taken into account through a self-consistently determined effective external field.
The theory contains an adjustable length 
parameter $\sigma$ that specifies the cut-off distance for the short-ranged interaction.
This can be chosen to minimize the errors resulting from the mean-field treatment of the 
complementary long-ranged part. Here we suggest that in many cases an accurate approximation to the
effective field can be obtained directly from
the equilibrium charge density given by the Debye theory of screening,
thus eliminating the need for a self-consistent treatment.
In the limit $\sigma\to0$, this assumption reduces to the
classical Debye approximation. We examine the numerical performance of this approximation for a
simple model of a symmetric ionic mixture. Our results for thermodynamic and structural properties
of uniform ionic mixtures agree well with similar results of Ewald simulations of the full ionic system.
In addition we have used the simplified theory in a grand-canonical simulation of a nonuniform
ionic mixture where an ion has been fixed at the origin. Simulations using short-ranged
truncations of the Coulomb interactions alone do not satisfy the exact condition of complete
screening of the fixed ion, but this condition is recovered when the effective field is taken into account.
We argue that this simplified approach can also be used in the simulations of more complex nonuniform systems.
\end{abstract}

\maketitle

\section {Introduction}

The long-ranged nature of the Coulomb interaction often causes problems in computer 
simulations.~\cite{HansenMcDonald} Although in some cases direct truncation of Coulomb 
interactions, e.g., by reaction field methods~\cite{ChargedCloud} or shifted force 
truncation~\cite{Wolf} can give accurate results,~\cite{Fennell} such methods have suffered from 
significant errors in many other physically relevant
cases.~\cite{ElectEffectsReview,bergdorf:9129,SchreiberH} Truncation 
schemes tend to work best in dense uniform systems, where there is considerable cancellation of 
the long-ranged electrostatic forces, and they often perform poorly in inhomogeneous 
systems.\cite{ElectEffectsReview,bergdorf:9129,SchreiberH,LMFPairingAndWalls,WeeksLMFSim}
At the same time, the Ewald sum method~\cite{Ewald,MeshEwald} --- which does not truncate
Coulomb interactions and accurately accounts for all images generated by periodic boundary
conditions --- is generally complex and computationally demanding when applied to inhomogeneous
systems,~\cite{Widmann} although considerable simplification is possible in reduced geometries such as 
a two-dimensional slab.~\cite{Shelley&Patey,Spohr,EW3DC,Holm1,Holm2} Finally, the strict periodicity of 
the Ewald sum method has been known to introduce artifacts in some biologically relevant studies.~\cite{EwaldArtifacts}

In this paper, we examine the performance of an alternative treatment of inhomogeneous Coulomb systems, 
local molecular field (LMF) theory,~\cite{WeeksLMFReview} when applied to a simple model for an ionic solution.
This theory accounts for the averaged effects of slowly-varying long-ranged components of the intermolecular 
interactions by using a self-consistently determined effective field. LMF theory provides a general framework 
for treating both uniform and nonuniform systems. We show for the ionic system considered in this paper that 
analytic results from the Debye theory of screening can be used to greatly simplify the theory and find very good 
agreement with results of conventional simulations using Ewald sums.

The derivation of LMF theory and its application to general Coulombic systems are given in detail
elsewhere.~\cite{WeeksLMFSim,WeeksYBG,WeeksYBG2,WeeksLMF,LMFCoulomb,LMFPairingAndWalls}
The ideas behind the theory are best understood by considering a nonuniform one component 
system with long-ranged intermolecular interactions $w(r)$ in an external field $\phi(\mathbf{r})$,
which can represent the interactions with fixed objects such as walls or solutes.
LMF theory relates structural and 
thermodynamic properties of the original system to those of a ``mimic system" with short-ranged 
interactions $u_0(r)$ in a renormalized effective field $\phi_R(\mathbf{r})$ that accounts for the 
averaged effects of the remaining long-ranged component $u_1(r)\equiv w(r)-u_0(r)$ of the 
intermolecular interactions. A key idea in LMF theory is that $u_1(r)$ should be
properly chosen to be slowly varying so that the averaging can give accurate results.

For such a $u_1(r)$, the effective field is determined in principle by the condition that the 
nonuniform singlet density in the mimic system (denoted by the subscript R) equals that in the 
original system:
\begin{equation}
\rho_R(\mathbf{r};[\phi_R])=\rho(\mathbf{r};[\phi]).
\label{eqn:LMFstructure}
\end{equation}
An explicit equation for $\phi_R(\mathbf{r})$ can be derived by subtracting the 
first equations of the exact Yvon-Born-Green hierarchy that relate the gradient of the singlet density to 
forces in the full and mimic systems. As argued in 
Refs.~[\onlinecite{WeeksLMFSim,WeeksYBG,WeeksYBG2,WeeksLMF,LMFCoulomb,LMFPairingAndWalls}],
when $u_1$ is chosen to be slowly varying over the range of  pair correlations between 
neighboring molecules, the effective field $\phi_R(\mathbf{r})$ is accurately given by the 
self-consistent LMF equation,
\begin{equation}
\phi_R(\mathbf{r})=\phi (\mathbf{r})+\int d{\mathbf{r}}^{\prime }\rho_R(\mathbf{r}^{\prime}; 
[\phi_R]) u_1(|\mathbf{r}^{\prime }-\mathbf{r}|) +C,
\label{eqn:LMFgeneral}
\end{equation}
where $C$ is a constant of integration. Equation~(\ref{eqn:LMFstructure}) then relates structure in 
the original and mimic systems; thermodynamic properties can be similarly related by integration 
over the structure. Finally, uniform systems are treated in LMF theory by 
choosing $\phi(\mathbf{r})$ to be the field arising from a fixed fluid particle, i.e., by taking 
$\phi(\mathbf{r})=w(r)$. This field will induce a nonuniform singlet density that can be directly 
related to the pair distribution function in the uniform fluid.~\cite{Percus}

LMF theory has many ideas in common with earlier methods, but they are implemented in new 
ways that avoid most of the limitations of those methods. As can be seen from 
Eq.~(\ref{eqn:LMFgeneral}), LMF theory uses a mean-field average of the long-ranged 
interactions, similarly in spirit to the random-phase approximation of density functional 
theory~\cite{DFTreview} or to the Debye theory of screening for ionic 
systems.~\cite{HansenMcDonald} But, crucially, the average in LMF theory is taken only over 
particular slowly varying components $u_1$, chosen precisely so that their averaged effects can 
be accurately described by an effective field.

The general idea of separating intermolecular interactions into short-ranged and long-ranged parts 
has also proved useful in many different contexts, particularly for uniform systems where 
long-ranged forces largely cancel and thus can be treated as a weak perturbation or by 
standard integral equation 
closures.~\cite{HansenMcDonald,Widom,WCA,KUH,LebStellBaer,DFTreview} For historical 
reasons the intermolecular potential $w$ was often split so that the properties of the reference 
system with the intermolecular potential $u_0$ would be well-known, e.g., a hard sphere fluid or 
an ideal gas. However, these choices do not generally guarantee that the remaining interactions 
$u_1$ can in fact be accurately treated by perturbation or diagrammatic methods.

With present day computers, complex short-ranged systems can be simulated efficiently and so the 
nature of the reference system is no longer an overriding issue. Therefore the split of the 
intermolecular potential $w$ in LMF theory can be optimized to minimize the errors associated 
with the approximate mean-field treatment of its long-ranged part $u_1$. The mimic system is then 
defined as the special short-ranged reference system resulting from this optimal split of $w$,
and we can use the simulations to accurately determine its properties. 

LMF theory thus corrects the two major shortcomings of the classical Debye theory of ionic systems, 
namely, its inaccurate mean-field averaging of the entire Coulomb potential and the highly 
approximate Boltzmann form of the density response to the effective field.
But as discussed below the (linearized) Debye theory
satisfies the exact Stillinger-Lovett moment conditions~\cite{SL} and therefore
correctly describes the asymptotic behavior of the charge correlation function.
These features of the Debye theory can be exploited to greatly
simplify the determination of the effective field in LMF theory while still giving accurate results, as
we now show.

\section {Application of LMF theory to a symmetric ionic fluid model}
We consider a uniform mixture of $N$ positive and $N$ negative ions with charges $+q$ and $-q$.
The molecular cores are described by the repulsive part of the LJ potential 
$u_{0}^{LJ}(r)$.\cite{WCA} 
The total intermolecular potential between ions with charges $q_i$ and $q_j$ ($=\pm q$) is taken 
to be
\begin{equation}
w_{ij}(r)=u_{0}^{LJ}(r)+\frac{q_{i}q_{j}}{r},  \label{wtotij}
\end{equation}
where $r$ is the distance between the ion centers.

The strength of the Coulomb interactions can be characterized by the ratio $\Gamma=l_B/d$ of the 
Bjerrum length $l_B=q^2/k_BT$ (the distance where the Coulomb energy between two positive 
ions equals $k_{B}T$) to the ``ion diameter" $d \equiv \sigma^{LJ}$, where $\sigma^{LJ}$ is the 
length parameter in the repulsive LJ potential. $\Gamma \gtrsim 1$
characterizes the ``strong coupling" regime.

\begin{figure}
\rotatebox{90}{\includegraphics[height=7cm,clip]{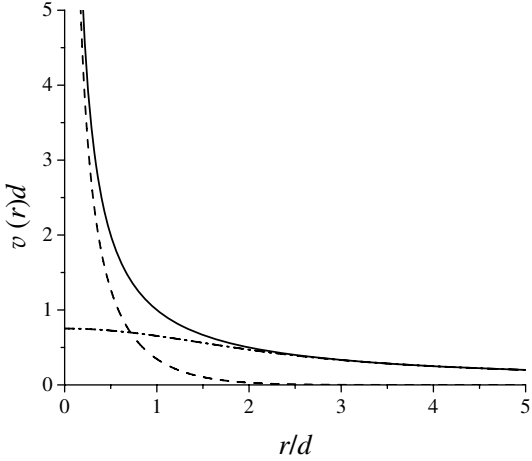}}
\caption{\small Separation of the Coulomb potential $v(r)=r^{-1}$ (solid) into
a short-ranged part $v_0(r,\sigma)= r^{-1}\mathrm{erfc}(r / \sigma)$ (dash)
and a long-ranged part $v_1(r,\sigma)= r^{-1}\mathrm{erf}(r / \sigma)$ (dash-dot).  The cut-off length $\sigma=1.5d$
shown here is also used in Fig.\ \ref{ghighdens}, as discussed below in Sec.\ III.}
\label{Split}
\end{figure}

Previous studies \cite{Ceperley,WeeksLMFSim,LMFCoulomb,LMFPairingAndWalls} have
shown that it is advantageous to divide the Coulomb interaction
$v(r)\equiv 1/r$ into short-ranged and long-ranged parts in the following way (see Fig.~\ref{Split}): 
\begin {equation}
\frac{1}{r}=\frac{\mathrm{erfc}(r / \sigma)}{r} + \frac{\mathrm{erf}(r / \sigma)}{r}
\equiv v_0(r,\sigma)+v_1(r,\sigma),
\label{erfc}
\end{equation}
where $\sigma$ is a length scale at our disposal. The idea behind this separation is best understood 
by considering the Fourier transform of the long-ranged component $v_1(r,\sigma)=\mathrm{erf}(r 
/ \sigma)/r$,
\begin{equation}
\hat{v}_1(k,\sigma)= \frac{4\pi}{k^2}\exp\left(-\frac{k^2\sigma^2}{4}\right).
\label{u1_fourier}
\end{equation}
The function $\hat{v}_1(k,\sigma)$ differs significantly from zero only at small wave-vectors 
$k\sigma \lesssim 2$. As a result, $v_1(r,\sigma)$ remains finite as $r\to0$ and is slowly varying for 
$r\lesssim\sigma$, while still decaying asymptotically as $1/r$. This makes it much more suitable
for a mean-field averaging than the full Coulomb  potential $v(r)$ used in the classical Debye theory.

This generates a separation of the intermolecular potentials $w_{ij}(r)\equiv u_{0,ij}(r) + 
u_{1,ij}(r)$ 
into short-ranged and long-ranged components respectively, where
\begin{equation}
u_{0,ij}(r)=u_{0}^{LJ}(r)+q_{i}q_{j}v_0(r,\sigma) \label{u0ij}
\end{equation}
and
\begin{equation}
u_{1,ij}(r)=q_{i}q_{j}v_1(r,\sigma). \label{u1ij}
\end{equation}

As mentioned above, LMF theory for a uniform system focuses on the density response to a fixed 
fluid particle. For the symmetric ionic system considered here, we can assume without loss of 
generality that a positive ion is fixed at the origin. This yields a single particle field 
$\phi_{j}(\mathbf{r}) = w_{+j}(r)$ acting on an ion with charge $q_j$. Using Eqs.\ (\ref{u0ij}) 
and (\ref{u1ij}), $\phi_{j}(\mathbf{r})$ naturally separates into a short-ranged core part 
$u_{0,+j}(r)$ and the long-ranged remainder $u_{1,+j}(r)$. 

The fixed ion induces a nonuniform singlet density $\rho_{j}(\mathbf{r})=\rho g_{+j}(r)$, 
proportional to the radial distribution function $g_{+j}(r)$ in the uniform fluid.\cite{Percus}
Here $\rho=N/V$ is the number density of positive or negative ions. Similarly, the induced charge 
density $\rho^{q}(\mathbf{r})\equiv q\rho_{+}(\mathbf{r})- q\rho_{-}(\mathbf{r})$ satisfies
\begin{equation}
\rho^{q}(\mathbf{r})=q\rho\left[g_{++}(r)-g_{+-}(r)\right]\equiv q\rho h^q(r).
\label{rho_g}
\end{equation}
The electrostatic energy per ion in the uniform ionic fluid can be written exactly
in terms of the charge density as
\begin{equation}
\frac{U}{2N}=\frac{q}{2}\int d{\mathbf r^{\prime}}\rho^{q}({\mathbf 
r^{\prime}})v(r^{\prime}) .
\label{energy_r_convolution}
\end{equation} 

LMF theory models the original uniform system by a nonuniform mimic system comprised of 
``solvent" mimic ions with short-ranged intermolecular interactions $u_{0,ij}(r)$ in an effective 
field $\phi_{R,j}(\mathbf{r})$, which we can picture as arising from a modified ``solute"
ion fixed at the origin. According to Eq.\ (\ref{eqn:LMFstructure}), when LMF theory is accurate, 
the solute-induced densities in the mimic system should equal the analogous densities in the 
original system. Using Eqs.\ (\ref{u0ij}) and (\ref{u1ij}), the LMF equation for
$\phi_{R,j}(\mathbf{r})$ can be written as [cf. Eq.\ (\ref{eqn:LMFgeneral})]
\begin{eqnarray}
&&\phi_{R,j}({\mathbf r})=u_{0,+j}(r) +\nonumber \\
&&~~~q_j\int d{\mathbf r^{\prime}}\left[q\delta({\mathbf r^{\prime}})+\rho^{q}_R ({\mathbf 
r^{\prime}})\right]
v_1(|{\mathbf r}-{\mathbf r^{\prime}}|,\sigma),
\label{field}
\end{eqnarray}
where $\rho^{q}_R({\mathbf r})=q\rho_{R,+}(\mathbf{r})- q\rho_{R,-}(\mathbf{r})$
is the induced charge density in the nonuniform mimic system.

The form of the effective field in Eq.\ (\ref{field}) clearly depends on $\sigma$. There are two 
criteria that help determine an optimal choice of $\sigma$. For the LMF 
method to be quantitatively valid, $\sigma$ should be large enough that $v_{1}(r, \sigma )$ 
remains slowly varying on the scale of short-ranged pair correlations. At the same time, it is 
desirable to keep $\sigma$ small in order to reduce simulation times of the mimic system. Thus 
$\sigma$ is generally chosen near its minimal accurate value $\sigma_{\mathrm{min}}$, which is 
state dependent and of the order of a characteristic neighbor
spacing.\cite{LMFPairingAndWalls,WeeksLMFSim,LMFCoulomb}

In principle, Eq.\ (\ref{field}) has to be solved self-consistently since the effective field 
$\phi_{R,j}(\mathbf{r})$ depends on the charge density $\rho^{q}_R(\mathbf{r})$ in the
presence of the field itself. During an iterative solution of Eq.\ (\ref{field}), the density induced by 
a given field can be accurately determined from the simulation of the nonuniform mimic system. 
This procedure has been successfully carried out for models of 
ions~\cite{WeeksLMFSim} or water~\cite{WaterUnpublished}
confined between charged and uncharged hard walls. Once the self-consistent charge 
density $\rho^{q}_R(\mathbf{r})$ has been determined, it can be used in 
Eq.\ (\ref{energy_r_convolution}) to calculate the electrostatic energy of the original ionic system. 

However, self-consistent simulations of the nonuniform mimic system can be computationally
demanding and may not always be required to obtain accurate results. In the following we 
introduce a simplified version of LMF theory that avoids the need for full self-consistency. We show that
both the local structure and the electrostatic energy of the uniform ionic system can be 
accurately reproduced by combining results of straightforward simulations of the \emph{uniform} 
mimic system with $\phi_{R,j}(\mathbf{r})=0$ along with analytic results from the Debye 
theory of screening for corrections arising from the long-ranged forces.
Similar ideas have also proved useful in simulations of more complex systems 
of charged polymers, as will be described elsewhere.~\cite{PolymerUnpublished}

\begin{figure}
\subfigure[$\Gamma=5$, $\rho d^3=0.3816$]
{
\rotatebox{90}{\includegraphics[height=7cm,clip]{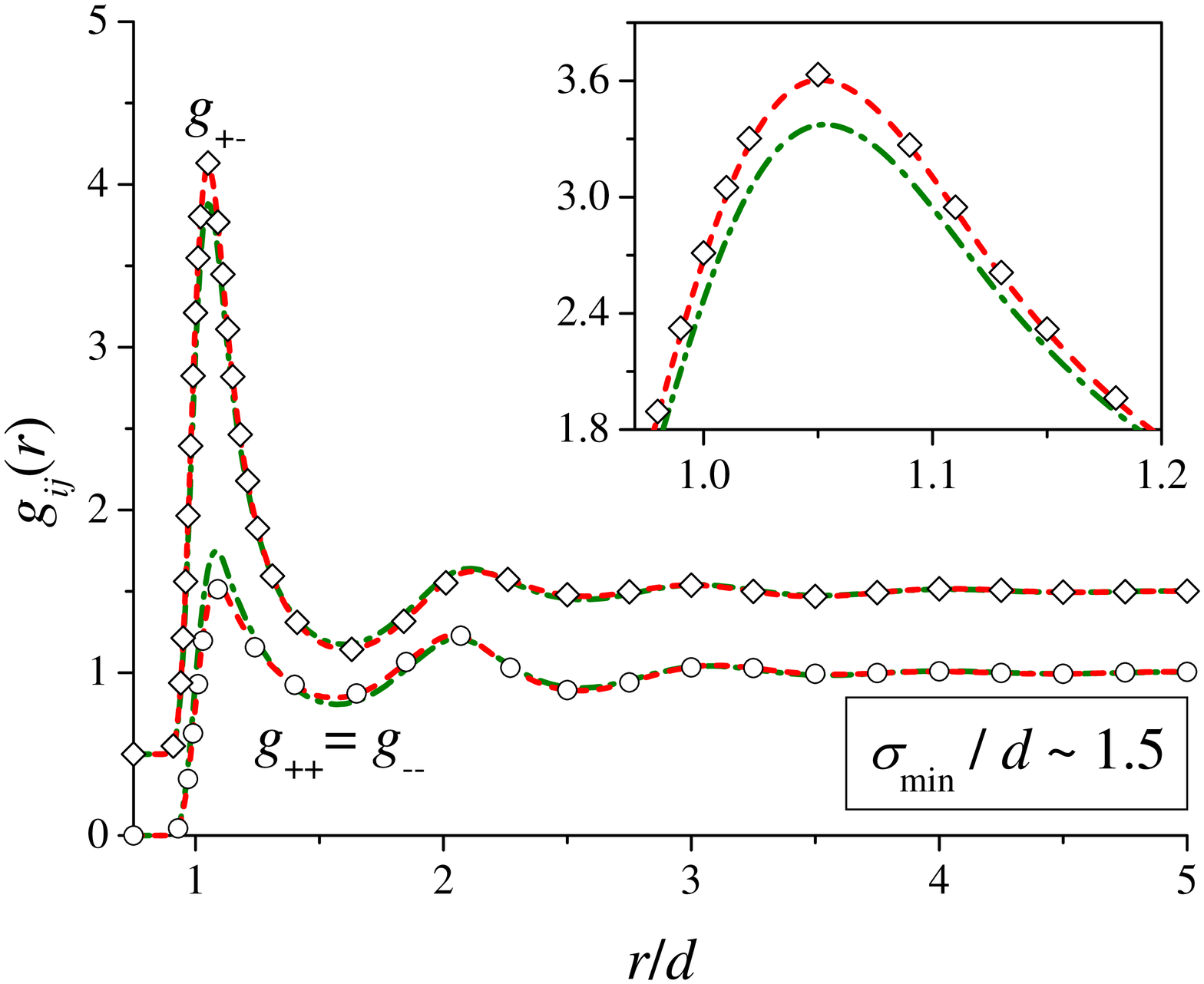}}
\label{ghighdens}
}\\
\subfigure[$\Gamma=5$, $\rho d^3=0.0012$]
{
\rotatebox{90}{\includegraphics[height=7cm,clip]{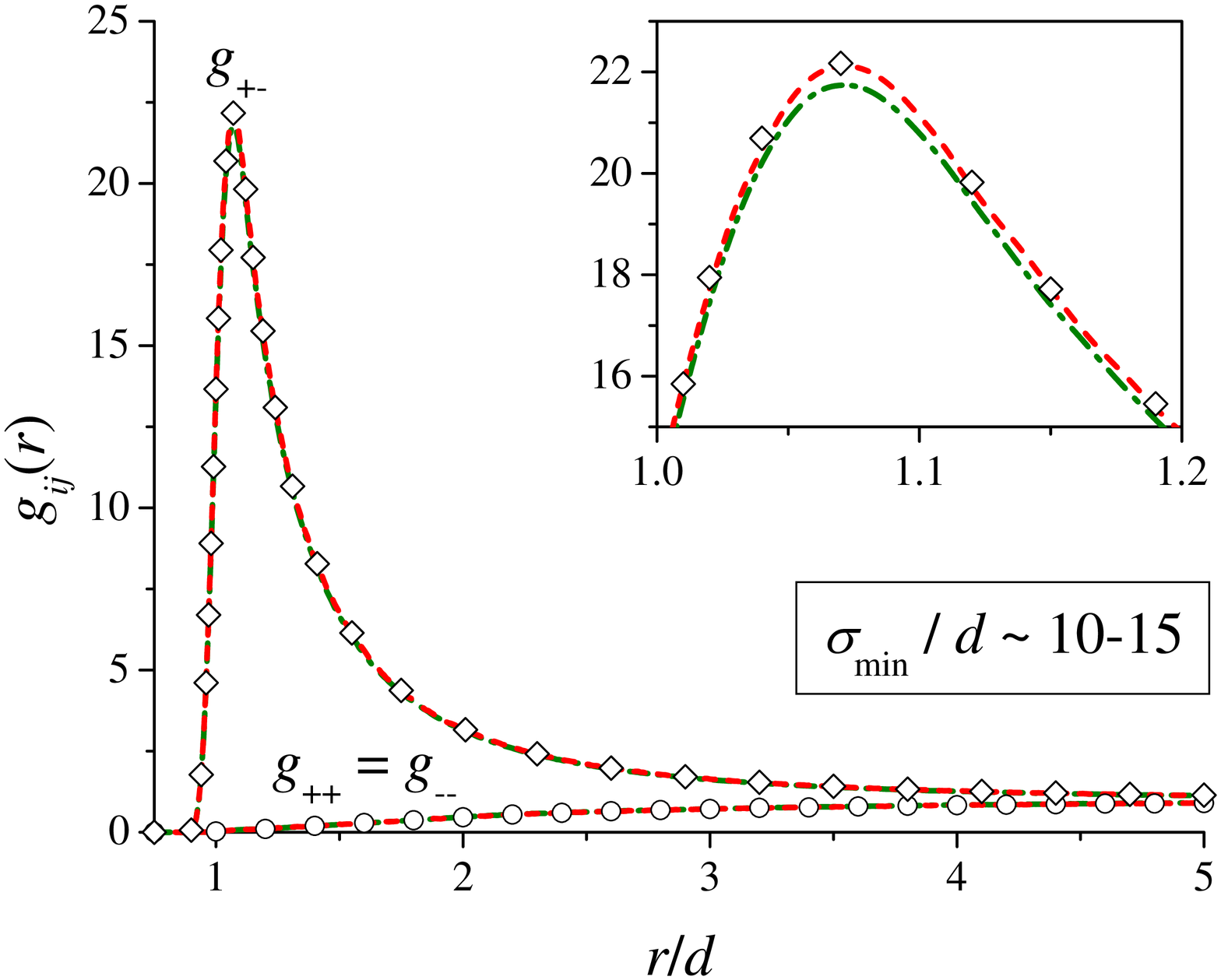}}
\label{glowdens}
}
\caption{\small The ion-ion distribution functions $g_{ij}(r)$ in two uniform symmetric ionic 
mixtures with $\Gamma=5$, $\rho d^3=0.3816$ \subref{ghighdens}, and $\Gamma=5$, $\rho 
d^3=0.0012$ \subref{glowdens}. The open symbols are the ion pair distribution functions in the 
full Coulomb systems, obtained using Ewald simulations. The curves are the same functions in the 
mimic systems with \subref{ghighdens} $\sigma/d =1.5$ (dash) and 1 (dash-dot), 
\subref{glowdens} $\sigma/d= 15$ (dash) and 10 (dash-dot). The $g_{+-}(r)$ functions in
(a) are vertically displaced by 0.5. The insets show a zoom of the first peak of $g_{+-}(r)$. 
}
\label{pcfs}
\end{figure}

\section {Local structure from short-ranged simulations}

\subsection {Strong coupling approximation}

At high density in the uniform ionic system we expect considerable
cancellation of the forces from the slowly-varying
part $v_1(r,\sigma)$ of the Coulomb interaction when $\sigma$ is chosen properly.
In the simplest ``strong-coupling approximation" (SCA) to the full LMF 
theory,\cite{LMFCoulomb,LMFPairingAndWalls,WeeksLMFSim}
we ignore all effects of $v_1(r,\sigma)$ on the fluid structure, i.e., neglect the integral 
in Eq.~(\ref{field}). Note that the strong short-ranged part $v_{0}(r,\sigma)$ of the Coulomb 
interaction as well as the LJ core is still taken into account in the SCA. This part of the 
interaction would be expected to dominate local structural arrangements at lower densities as well, 
provided that $\sigma$ is chosen large enough.

In the SCA the self-consistent field $\phi_{R,j}(\mathbf{r})$ in Eq.~(\ref{field}) is approximated 
by $\phi_{0,j}(\mathbf{r})\equiv u_{0,+j}(r)$, the known field of a solvent mimic ion.
In this case, the induced charge 
density $\rho^{q}_0({\mathbf r})$ can be determined directly and
more efficiently from the radial distribution 
functions $g_{0,+j}(r)$ in the \emph{uniform} mimic system [cf. Eq.~(\ref{rho_g})],
\begin{equation}
\rho_0^q(\mathbf{r})=q\rho\left[g_{0,++}(r)-g_{0,+-}(r)\right]\equiv q\rho h_0^q(r).
\label{rho_g_0}
\end{equation}
Thus the SCA can be viewed as a particularly useful direct truncation scheme,~\cite{Wolf} whose 
accuracy can be justified in certain limits and corrected, if necessary, by the full LMF theory.
Note that we use the subscript $0$ to refer both to pair correlation functions in
the uniform mimic system with $\phi_{R,j} =0$ and to the equivalent singlet densities in the
nonuniform system when $\phi_{R,j}$ is approximated using the SCA by $\phi_{0,j}$.
The subscript $R$ refers to the nonuniform mimic system in the presence of the full renormalized
field given by Eq.~(\ref{field}).

Figure~\ref{pcfs} gives representative results of Langevin dynamics simulations
using the Ewald sum method for the present ionic fluid model at states with fixed
temperature $k_{B}T/\epsilon^{LJ}=1$ and charge $q$ chosen so that the system is at 
moderately strong coupling with $\Gamma  = 5$. These are compared to simulations of
the uniform mimic system, as described by the SCA, where the cut-off radius for
the short-ranged potential $v_0(r,\sigma)$ is $2.5\sigma$. 

Figure \ref{ghighdens} shows results for a high density state with $\rho d^{3}=0.3816$,
where there is substantial cancellation of attractive forces. We find that for $\sigma /d = 1.5$
there is excellent agreement between the distribution functions in the full Coulomb and 
mimic systems over the range of $r$ shown. Higher values of $\sigma$
give equally good results, but smaller $\sigma$ values cause 
noticeable errors, as illustrated in the inset for $\sigma /d =1$, so $\sigma_{\mathrm{min}}$
is about $1.5d$ for this state. This good agreement is consistent with previous 
work on LMF theory~\cite{WeeksLMFSim,LMFCoulomb,LMFPairingAndWalls}
and with earlier findings that models with
truncated Coulomb interactions can often provide a good description of structural
features in dense uniform systems.~\cite{Wolf,Fennell}

Figure \ref{glowdens} makes the same comparison for a very low density state with $\rho 
d^{3}=0.0012$. Despite this low density the coupling is strong enough that the Debye theory alone
would give poor results, and this state presents a major challenge to theory.
We find that a much larger value of $\sigma/d =15$ is needed to achieve 
comparably good results (the small deviations for $\sigma/d =10$ are shown in the inset). Large 
$\sigma$ is needed since there is essentially no force cancellation at low densities and the 
characteristic neighbor distances are large, but this also makes simulations of the mimic system 
much more costly in this regime.

\begin{figure}
\subfigure[$\Gamma=5$, $\rho d^3=0.3816$]
{
\rotatebox{90}{\includegraphics[height=7cm,clip]{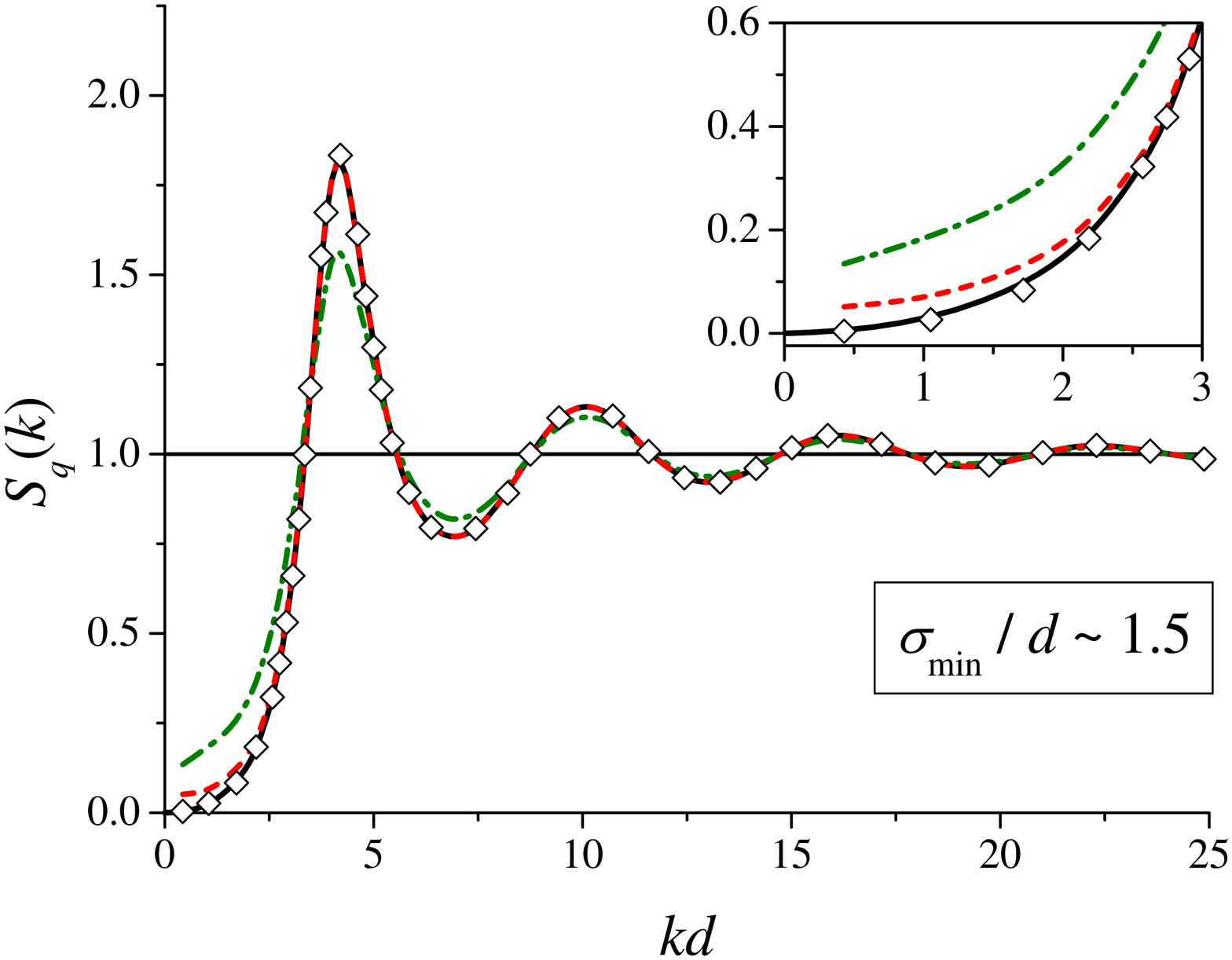}}
\label{Shighdens}
}\\
\subfigure[$\Gamma=5$, $\rho d^3=0.0012$]
{
\rotatebox{90}{\includegraphics[height=7cm,clip]{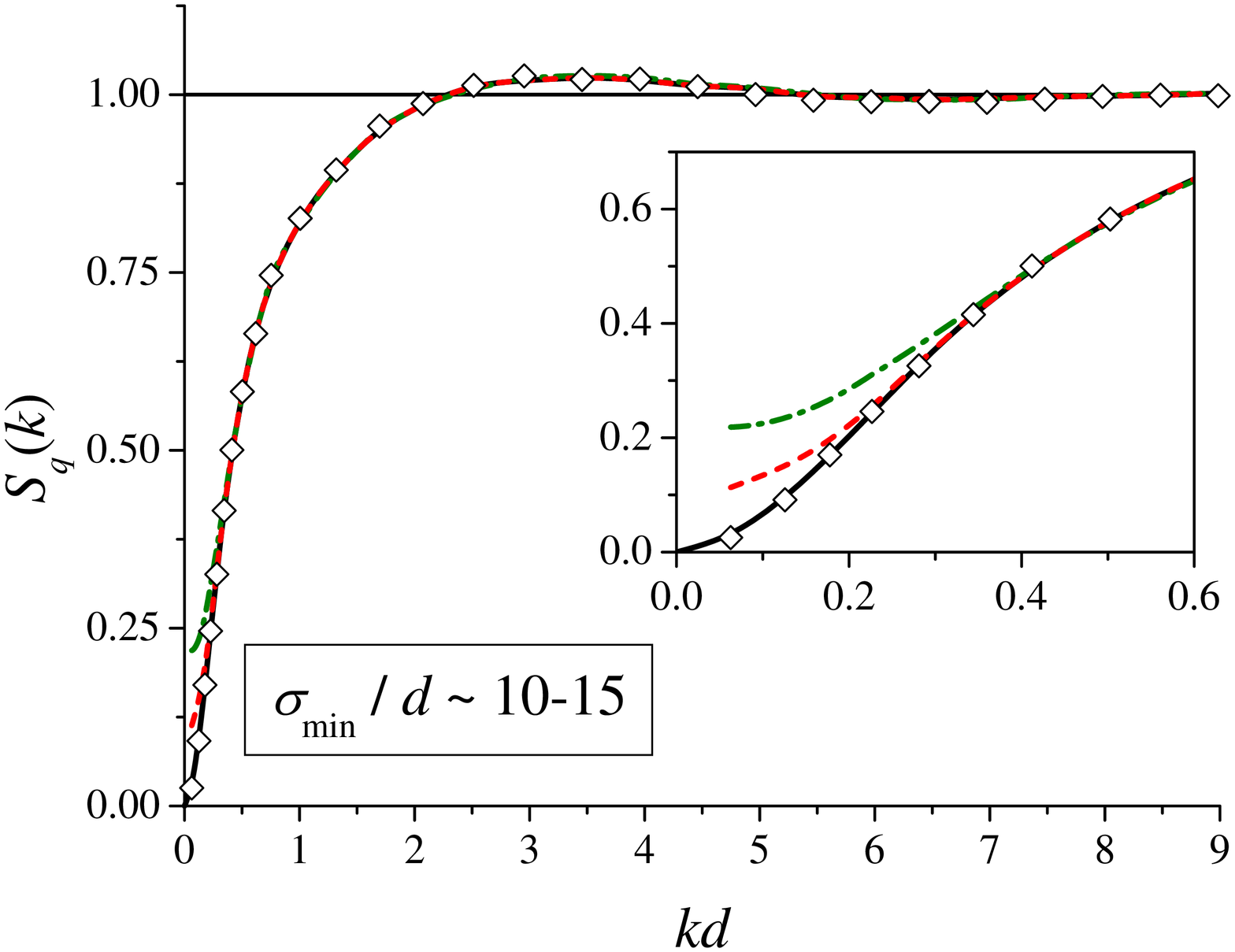}}
\label{Slowdens}
}
\caption{\small The charge structure factors $S^q(k)$ for the same systems as in Fig.~\ref{pcfs}. 
The insets show a zoom of the small-$k$ region where a discrepancy is seen in the structure factors 
of the full Coulomb and mimic systems. The solid curves show the results of application of the 
perturbation equation~(\ref{RPA1_structs}) to the dashed curves in Figs.~\ref{Shighdens} and 
\ref{Slowdens}.}
\label{structs}
\end{figure}

However, despite the excellent agreement between the distribution functions in the original and 
uniform mimic systems in the range of $r$ shown, there are fundamental differences in the
asymptotic behavior of these functions as $r\to\infty$. This is most easily seen from the small-$k$ 
behavior of the charge structure factor $S^q(k)$, which is simply related to the Fourier transform of 
the dimensionless charge correlation function $h^q(r)$ defined in Eq.~(\ref{rho_g}),
\begin{equation} 
S^q(k)=1+\rho\hat{h}^q(k).
\label{structure_factor}
\end{equation}
As $k\to0$, the charge structure factor $S^q(k)$ of any ionic system exhibits the same universal 
behavior,
\begin{equation}
S^q(k)=k^2\lambda_D^2+O(k^4),
\label{SL}
\end{equation}
where 
\begin{equation}
\lambda_D= (8\pi l_B\rho)^{-1/2}
\label{debyelength}
\end{equation}
is the Debye screening length. The exact form in Eq.~(\ref{SL}) is independent of any details of 
the short-ranged core interactions $u_{0,ij}(r)$ and is a consequence of the Stillinger-Lovett 
moment conditions.~\cite{SL} In contrast, the analogous function
\begin{equation} 
S_0^q(k)=1+\rho\hat{h}_0^q(k)
\label{structure_factor_0}
\end{equation}
for the uniform short-ranged mimic system, where $h_0^q(r)$ is defined in Eq.~(\ref{rho_g_0}), 
will 
remain finite as $k\to0$, with the coefficient of $k^2$ depending on the details of the 
intermolecular interactions.

In Fig.~\ref{structs} we compare $S^q_0(k)$ and $S^q(k)$ for the same ionic mixtures whose  
$g_{ij}(r)$ are shown in Fig.~\ref{pcfs}. For $\sigma\gtrsim\sigma_{\mathrm{min}}$, the 
functions 
$S^q_0(k)$ closely follow $S^q(k)$ everywhere except for small $k$, where the differences 
described above can be seen. If $\sigma<\sigma_{\mathrm{min}}$, noticeable discrepancies 
between 
$S^q_0(k)$ and $S^q(k)$  appear also at larger $k$, as illustrated by the dash-dot curve in 
Fig.\ \ref{Shighdens}.

\subsection {\label{nonuniform} Complete screening and the Debye theory}

This different behavior at small wave-vectors implies that
the uniform mimic system will not exactly satisfy the basic ``complete screening condition"
that true ionic fluids obey. Complete screening (equivalent to the Stillinger-Lovett
zeroth moment condition) requires that the exact $\rho^q$ induced by
a fixed positive ion in a grand ensemble will satisfy
\begin{equation}
\int d{\mathbf r}\rho^q({\mathbf r})= -q.
\label{full neutrality}
\end{equation}
However, the results above for the simpler SCA imply
that Eq.\ (\ref{full neutrality}) will not hold if $\rho^q$ is approximated by the $\rho_0^q$
given by a grand canonical simulation of the mimic system with a fixed
positive ``solvent'' mimic ion at the origin.

To verify these conclusions, we have performed grand-canonical ensemble simulations of the 
nonuniform mimic system in the field $\phi_{0,j}(\mathbf{r})=u_{0,+j}(r)$ for a state with 
$\Gamma = 5$, $\rho d^3 = 0.0012$ and $\sigma = 10d$. We find that the  mean number of 
counterions $\langle N \rangle = 1202.19$ in the simulation box exceeds the mean number of 
coions by $\Delta N=0.68\pm 0.10$. This is definitely smaller than the mean difference $\Delta N 
=1$ that should hold in the case of complete screening. We chose a very large simulation box with 
$L=100d$ such that this value of $\Delta N$ is independent of $L$ and carried out a long 50ns 
simulation run to obtain the reported statistical convergence of $\Delta N$.

In addition, we ran a similar simulation of the nonuniform mimic system in which the long-ranged 
part $v_1(r,\sigma)$ of the Coulomb potential has been taken into account through the LMF 
equation~(\ref{field}). One can show that the density induced by a self-consistent solution
of this equation will exactly satisfy the complete screening condition.\cite{LMFCoulomb}
However, instead of solving Eq.~(\ref{field}) self-consistently, we have replaced 
the charge density profile $\rho_R^q(r)$ in this equation by the
screening profile of a point charge given by the linearized Debye-H\"{u}ckel theory: 
\begin{equation}
\rho_D^q(r)=-\frac{q}{4\pi\lambda_D^2r}\exp\left(-\frac{r}{\lambda_D}\right).
\label{Debye_profile}
\end{equation}

At first glance this may seem to be a very crude
approximation, since the Debye profile is generally accurate only when
both $\Gamma$ and $\rho$ are very small. Otherwise $\rho_R^q(r)$ and $\rho_D^q(r)$
will differ considerably at small $r$. However,
the Debye profile has the correct asymptotic behavior since it
satisfies the exact Stillinger-Lovett
zeroth and second moment conditions.~\cite{SL} Moreover, when integrated over the slowly varying part of 
the Coulomb potential $v_1(r,\sigma)$ as in Eq.\ (\ref{field}), most of the short-ranged features
of this profile on the scale $r\lesssim\sigma$ become irrelevant,
so the resulting estimate for $\phi_{R,j}({\mathbf r})$ can still be accurate.

With this approximation, the integration in Eq.~(\ref{field}) can be carried out exactly and
we obtain an explicit expression for the effective field
$\phi_{R,j}({\mathbf r})$,
\begin{eqnarray}
\phi_{R,j}({\mathbf r})\approx u_{0,+j}(r)+\frac{q_jq}{2r}\exp\left(\frac{\sigma^2}{4\lambda_D^2} 
\right) \nonumber\\
\times\left[\exp\left(-\frac{r}{\lambda_D}\right)\mathrm{erfc}\left(\frac{\sigma}{2}-\frac{r} 
{\sigma}\right)\right. \nonumber\\
\left.-\exp\left(\frac{r}{\lambda_D}\right)\mathrm{erfc}\left(\frac{\sigma}{2}+\frac{r}{\sigma} 
\right) \right].
\label{Debye_field}
\end{eqnarray}

The simulation of the nonuniform mimic system, where $\phi_{R,j}({\mathbf r})$ is given by 
Eq.~(\ref{Debye_field}), yields $\Delta N=1.09\pm0.10$. Thus our simple
estimate for the effective field using the Debye theory can reproduce the complete screening 
behavior
seen in the full system or from a complete self-consistent solution of the LMF equation.

\section {Electrostatic energy}

Thermodynamic properties of ionic systems also require careful
attention to contributions from the long-ranged
parts of the Coulomb interactions. Again we find that analytic results from the Debye theory
can provide simple but accurate corrections to results from the uniform mimic system.
In analogy with Eq.~(\ref{energy_r_convolution}), the ``electrostatic energy" of the uniform 
mimic system is given by
\begin{equation}
\frac{U_0}{2N}=\frac{q}{2}\int d{\mathbf r^{\prime}}\rho^{q}_0({\mathbf r^{\prime}}) 
v_0(r^{\prime},\sigma).
\label{energy_r_convolutionmimic}
\end{equation} 
As illustrated in Table \ref{energy_table}, $U_0$ differs considerably from the full Coulomb 
energy $U$, determined by the Ewald sum method, even for 
$\sigma\gtrsim\sigma_{\mathrm{min}}$ when local structural properties of the original and mimic 
systems closely resemble each other.

\begin{table}
\begin{tabular}{c|c|c|c|c}
\hline\hline
\multirow{2}{*}{$\frac{\displaystyle\beta U}{\displaystyle 2N}$} & \multicolumn{2}{c}{$\rho 
d^3=0.0012$} & \multicolumn{2}{|c}{$\rho d^3=0.3816$} \\ \cline{2-5} 
& $\sigma=10$ & $\sigma=15$ & $\sigma=1$ & $\sigma=1.5$\\ \hline\hline
$\frac{\displaystyle\beta U_0}{\displaystyle 2N}$ & $-0.5946(4)$ & $-0.6837(3)$ 
& $-0.4912(1)$ & $-1.3122(2)$\\ \hline
Debye & $-0.8487(4)$ & $-0.8622(3)$ & $-3.2068(1)$ & $-3.1597(2)$\\
RPA& $-0.8453(5)$ & $-0.8612(3)$ & $-3.0136(2)$ & $-3.1335(2)$\\
Debye-M& $-0.8678(4)$ & $-0.8701(3)$ & $-3.2979(1)$ & $-3.1906(2)$ \\ 
\hline\hline 
$\frac{\displaystyle\beta U_{EW}}{\displaystyle 2N}$ & \multicolumn{2}{c}
{$-0.8708(4)$} & \multicolumn{2}{|c}{$-3.1880(3)$} \\
 \hline
\end{tabular}
\caption{\small The electrostatic energy $U$ for the same systems as in Figs.~\ref{pcfs} and 
\ref{structs}. The error in the last significant figure is indicated in parentheses. Approximations
for the total energy labeled Debye, RPA, and Debye-M are discussed in Eqs.\ (\ref{energy_correction_debye}),
(\ref{RPA1_energy}), and (\ref{RPA2_energy}) respectively.}
\label{energy_table}
\end{table}

To find the needed correction, we note that Eq.\ (\ref{energy_r_convolution}) can be exactly 
rewritten as
\begin{eqnarray}
\frac{\beta U}{2N}&=&\frac{\beta U_0}{2N}+\frac{l_B}{2}\frac{1}{(2\pi)^3}\int d\mathbf{k}
\left[S^q(k)-1\right] \hat{v}_1(k,\sigma)\nonumber \\
&+ &\frac{\rho l_B}{2}\int d{\mathbf r^{\prime}}[h^q(r^{\prime})-h_0^q(r^{\prime})] 
v_0(r^{\prime},\sigma),
\label{energy_r_convolution_split1}
\end{eqnarray}
where $h^q(r)$, $h_0^q(r)$, $S^q(k)$ and $U_0$ are defined in Eqs.~(\ref{rho_g}), 
(\ref{rho_g_0}), (\ref{structure_factor}) and (\ref{energy_r_convolutionmimic}), respectively.      

We expect that the value of the last integral in Eq.~(\ref{energy_r_convolution_split1}) is very 
small since with proper choice of $\sigma$ $h^q(r)$ and $h_0^q(r)$ are very similar over the entire
range of $r$ where $v_0(r,\sigma)$ differs significantly from zero (see Fig.\ \ref{pcfs}).
Hence the energy difference $\Delta U\equiv U-U_0$ can be accurately estimated as
\begin{equation}
\frac{\beta\Delta 
U}{2N}\approx\frac{l_B}{2}\frac{1}{(2\pi)^3}\int d\mathbf{k}\hat{v}_1(k,\sigma)\left[S^q(k)-
1\right].
\label{energy_k_convolution_split2}
\end{equation}
Note that the LMF integral in Eq.\ (\ref{field}) for $r=0$ equals $\Delta U/N$.

The function $\hat{v}_1(k,\sigma)$, given by Eq.\ (\ref{u1_fourier}), is a rapidly decaying 
function of $k$ for $k\sigma \gtrsim 2$. Thus only the small-$k$ behavior of $S^q(k)$ is 
significant in Eq.~(\ref{energy_k_convolution_split2}). 
Similar to our discussion of complete screening in Sec.~\ref{nonuniform}, this suggests
that we can accurately use the Debye approximation
\begin{equation}
S_D^q(k)=\frac{k^2}{k^2+\lambda_D^{-2}}.
\label{structs_debye}
\end{equation}
for the charge structure factor $S^q(k)$ in Eq.~(\ref{energy_k_convolution_split2}).
 
Equation~(\ref{structs_debye}) is exact at small enough $k$ since it
satisfies both Stillinger-Lovett moment conditions and, unlike 
Eq.~(\ref{SL}), correctly reduces to unity at large $k$. Furthermore it becomes an exact result for all $k$
in the limit of very small $\Gamma$ and $\rho$. Substituting Eq.\ (\ref{structs_debye}) in 
Eq.~(\ref{energy_k_convolution_split2}), we find
\begin{equation}
\frac{\beta\Delta U}{2N}=\frac{\beta U_D}{2N}f_1\left(\frac{\sigma}{\lambda_D}\right),
\label{energy_correction_debye}
\end{equation}
where $U_D$ is the well known result for the Coulomb energy in the Debye approximation,
\begin{equation}
\frac{\beta U_D}{2N}=-\frac{l_B}{2\lambda_D},
\label{energy_debye}
\end{equation}
and
\begin{equation}
f_1(y)= \exp \left( \frac{y^2}{4}\right) \mathrm{erfc}\left(\frac{y}{2}\right).
\label{f2}
\end{equation}

We expect accurate results from Eq.~(\ref{energy_correction_debye}) only when
$\sigma$ is properly chosen to be greater
than a state-dependent minimum value $\sigma_{\mathrm{min}}$.
For strong coupling states with $\Gamma \gtrsim 1$, we note that
$\sigma_{\mathrm{min}} \gg \lambda_D$. Using the asymptotic expansion of
$\mathrm{erfc}(y/2)$ in Eq.~(\ref{f2}), Eq.~(\ref{energy_correction_debye}) then reduces to the
strong coupling energy correction
\begin{equation}
\frac{\beta\Delta U}{2N}\approx -\frac{l_B}{\sqrt{\pi}\sigma}\left[1-
2\left(\frac{\lambda_D}{\sigma}\right)^2\right]
\label{Estrongcoupling}
\end{equation}
derived in Ref.~\onlinecite{LMFCoulomb}, where $S^q(k)$
was approximated by the second moment term in Eq.~(\ref{SL}).
But Eq.~(\ref{energy_correction_debye}) also correctly reduces to the exact Debye energy in
the limit of very weak coupling and low density where $\sigma_{\mathrm{min}} \to 0$
and provides a more generally useful expression.

\begin{figure}[t!b]
\subfigure[$\Gamma=5$]
{
\rotatebox{90}{\includegraphics[height=7cm,clip]{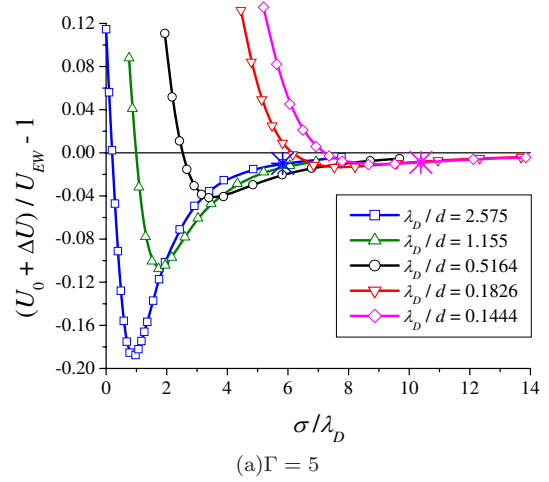}}
\label{a_correct}
}\\
\subfigure[$\rho d^3 = 0.02984$]
{
\rotatebox{90}{\includegraphics[height=7cm,clip]{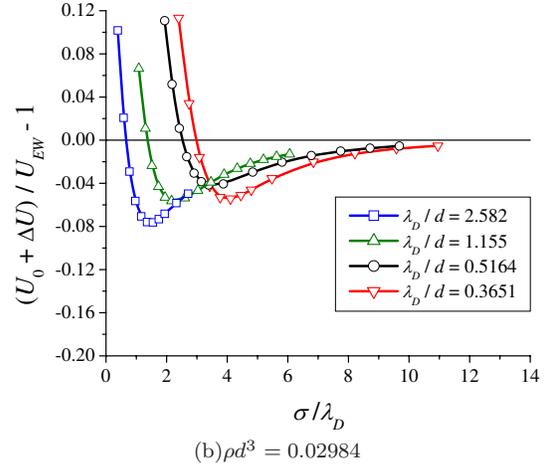}}
\label{b_correct}
}
\caption{\small The Debye corrected energies $U_0+\Delta U$ from
Eq.~(\ref{energy_correction_debye}) for a wide range of $\sigma$ relative to the 
Ewald energies $U_{EW}$ of the full Coulomb systems. The data sets, shown from left to right, 
are for \subref{a_correct} $\Gamma=5$ and $\rho d^3 = 0.0012$, 0.005969, 0.02984, 0.2387, 
0.3816, and for \subref{b_correct} $\rho d^3 = 0.02984$ and $\Gamma=0.2$, 1, 5, 10. The two 
stars in \subref{a_correct} show the location of $\sigma_{\mathrm{min}}$ as estimated from
the structure shown in Fig.\ \ref{pcfs} for a low density state with $\rho d^3 = 0.0012$ and
$\sigma_{\mathrm{min}}=15$,  and a high density state with $\rho d^3 = 0.3816$, and
$\sigma_{\mathrm{min}}=1.5$. A spline curve is fitted through each set of data to guide the eye.}
\label{correct_energies}
\end{figure}

The resulting energy estimates $U_0+\Delta U$ from Eq.~(\ref{energy_correction_debye}),
labeled ``Debye", are given in Table \ref{energy_table}
and are plotted for more states over a wide range of $\sigma$ in
Fig.~\ref{correct_energies}. We see that the 
deviations of the Debye corrected energy from the Ewald
energies $U_{EW}$ are reduced by approximately an 
order of magnitude from the uncorrected results given in Table \ref{energy_table}.
Moreover, the accuracy of the correction $\Delta U$ sharply increases with 
larger values of the ratio $\sigma/\lambda_D$. This allows us to determine
appropriate values for $\sigma_{\mathrm{min}}$ and obtain energy estimates accurate 
to within $1\%$ when $\sigma$ is chosen large enough.

In the limit $\sigma\to0$, $U_0$ becomes vanishingly small and Eq.~(\ref{energy_correction_debye}) 
reduces to the classical Debye approximation. Since the Debye theory generally overestimates the 
absolute value of $U$, all the curves in Fig.~\ref{correct_energies} turn up sharply at
small enough $\sigma < \sigma_{\mathrm{min}}$.

These results emphasize that LMF theory is an inherently approximate approach and accurate results
can be expected only when $\sigma$ is chosen greater than some state-dependent minimum value
$\sigma_{\mathrm{min}}$. In the present simplified treatment, errors for a wide range of $\sigma$ have been 
determined from comparison with the results of Ewald simulations. These fairly basic simulation studies can 
be used to devise an estimate of $\sigma_{\mathrm{min}}$ for an ionic solution of given concentration and ionic 
strength.

Although the energy correction in Eq.~(\ref{energy_correction_debye}) depends only on the ratio 
$\sigma/\lambda_D$, the numerical performance of this correction depends significantly on another length scale, 
the the characteristic neighbor distance between a pair of ions, estimated here by $r^*=(2\rho)^{-1/3}$.
As can be deduced from Fig.~\ref{correct_energies}, 
the accuracy of $\Delta U$ decreases sharply when $\sigma<r^*$. When $\sigma/r^*>1$, the accuracy of $\Delta U$ is 
generally acceptable and is higher for larger values of $\sigma/\lambda_D$. As a rule of thumb for systems with
moderately strong coupling, we suggest that if an ionic 
solution is simulated as part of a more complex system, the two conditions
$\sigma\gtrsim2r^*$ and $\sigma\gtrsim5\lambda_D$ should be 
fulfilled simultaneously to yield an accurate description of electrostatic interactions.

\section {Integral equation methods for a uniform ionic fluid}

Integral equation methods that treat the long-ranged part of intermolecular interactions as a weak 
perturbation\cite{KUH,LebStellBaer} have long been used to study ionic 
fluids,\cite{HansenMcDonald} and could serve as an alternative
approach to LMF theory for uniform systems.
These equations are usually derived by summing certain classes 
of diagrams, where individual diagrams represent different terms in the perturbation series for 
structural and thermodynamic quantities. Although it is difficult to develop a physical intuition for 
what the errors will be in a given application, these methods are expected to work best when the 
effects of the long-ranged perturbations on the structure of a short-ranged reference system are in 
some sense small.

Since the local structure is well described by the uniform mimic system,
it seems likely that this could serve as a particularly useful reference system
for perturbation integral equation methods. This idea was in fact 
suggested long ago by Ceperley and Chester,\cite{Ceperley} although they did not discuss the 
possibility of applying it outside the framework of integral equations. Here we use
one of the earliest perturbation approaches, the RPA-like method of 
Ref.~\onlinecite{KUH}, to correct results for the uniform mimic system. In this approach the charge structure 
factor $S^q(k)$ of a uniform ionic mixture can be approximately written as
\begin{equation}
S^q(k)=\frac{S_0^q(k)}{1+2l_B\rho\hat{v}_1(k,\sigma)S_0^q(k)},
\label{RPA1_structs}
\end{equation}
where $S_0^q(k)$ is the charge structure factor of the uniform mimic system and 
$\hat{v}_1(k,\sigma)$ is given in Eq.~(\ref{u1_fourier}). 

The resulting functions $S^q(k)$ are 
shown in Fig.~\ref{structs} where they are compared with $S_0^q(k)$ obtained from short-ranged 
simulations. In contrast with $S_0^q(k)$, the functions $S^q(k)$ satisfy both Stillinger-Lovett 
moment conditions. For $\sigma\gtrsim\sigma_{\mathrm{min}}$ they are virtually 
indistinguishable from the results of Ewald simulations. 

The perturbation method of Ref.~\onlinecite{KUH} also yields an expression for the energy 
correction $\Delta U$,
\begin{equation}
\frac{\beta\Delta U}{2N}=\frac{l_B}{2}\frac{1}{(2\pi)^3}\int \left[\frac{1}{2l_B\rho} 
\ln\frac{S^q_0(k)}{S^q(k)}-\hat{v}_1(k,\sigma)\right]d\mathbf{k},
\label{RPA1_energy}
\end{equation}
where $S^q(k)$ is given by Eq.~(\ref{RPA1_structs}). The energy estimates resulting from 
Eq.~(\ref{RPA1_energy}), labeled as ``RPA'' are given in Table~\ref{energy_table}.

Remarkably, although the calculation based on the Debye profile is much less involved, it gives 
energy estimates that are equally accurate. We find, however, that yet another energy correction 
gives even more accurate results,
\begin{equation}
\frac{\beta\Delta U}{2N}= \frac{\beta U_D}{2N}-\frac{\beta U_{0D}}{2N}.
\label{RPA2_energy}
\end{equation}
Here $U_{0D}$ is the ``electrostatic energy'' of the uniform mimic system
in the Debye limit,
obtained by summing for ``Coulomb cores" $v_0(r,\sigma)$ the same ring diagrams 
that lead to the conventional Debye expression $U_{D}$ when using the full Coulomb interaction $v(r)$.
This gives
\begin{equation}
\frac{\beta U_{0D}}{2N}=\frac{\beta U_D}{2N}f_3\left(\frac{\sigma}{\lambda_D}\right),
\label{reference_energy_debye}
\end{equation}
where
\begin{equation}
f_3(y)=\frac{2}{\pi}\int_0^{\infty}\frac{\displaystyle\left[1-\exp\left(-\frac{k^2y^2}{4}\right) 
\right]^2}{\displaystyle k^2+1-\exp\left(-\frac{k^2y^2}{4}\right)}dk.
\label{f3}
\end{equation}
 
The idea behind Eq.\ (\ref{RPA2_energy}) is that with a proper choice of $\sigma$,
the energy correction $\Delta U$ should be independent of most details of the short-ranged
interactions. Most errors in the Debye treatment of the short-ranged part of the Coulomb interactions are
canceled by subtraction of the two terms in Eq.\ (\ref{RPA2_energy}).
The results in Table \ref{energy_table}
for  this ``Debye-Mimic" (Debye-M) approximation  give best agreement
with the Ewald energies $U_{EW}$ of the full Coulomb systems. 
This seems to indicate that approximations of the RPA type work best if perturbation terms, similar 
to $\Delta U$, contain no information on the short-ranged core structure of the mimic system.

\section {Concluding remarks}

In this paper we have used a simplified version of local molecular field (LMF) theory~\cite{WeeksLMFReview} to 
calculate the structural and thermodynamic properties of a symmetric ionic mixture. LMF theory 
has already been applied successfully to the description of uniform and nonuniform ionic 
systems,\cite{LMFCoulomb,LMFPairingAndWalls,WeeksLMFSim} where the LMF was 
either approximated using only the SCA or was determined by a full self-consistent calculation.
We have shown for the ionic mixture considered in this paper that we can go beyond the SCA
but avoid the necessity of finding a self-consistent solution if we replace the mimic system's
charge density with its Debye analogue in the LMF equation~(\ref{field}). The resulting effective field,
given in Eq.~(\ref{Debye_field}), is sufficiently accurate to 
reproduce the exact complete screening condition in grand-canonical
ensemble simulations. Furthermore, the 
energies of uniform ionic mixtures, obtained under this assumption, agree well with those 
calculated using diagrammatic techniques and the Ewald sum method.

When applied to complex inhomogeneous systems, this assumption will significantly speed up the 
simulations of underlying mimic systems since it reduces the simulations in an \emph{a priori} 
unknown self-consistent field to the simulations in a simpler and well-defined external field.
In addition, the Debye charge density is found analytically from the Debye-H\"{u}ckel equation
for an infinite system. For unbounded Coulomb systems this should result in more accurate
values of the effective field in Eq.~(\ref{eqn:LMFgeneral}) than those obtained by numerical
integration using simulation results in a finite volume. 

These ideas are being actively applied in our studies of polyelectrolytes
in salt solutions.\cite{PolymerUnpublished}
Since the polymer dynamics are slow in comparison with the dynamics of small ions,
it is valid to assume that the distribution of salt ions around a polyelectrolyte is always
in local equilibrium. In this case, the total effective field of a polyelectrolyte is simply a sum
of the effective fields of all its monomer charges, given by Eq.~(\ref{Debye_field}).
This essentially amounts to replacing the full Coulomb potential of a monomer charge
with a screened Coulomb potential, defined as a sum of $r^{-1}\mathrm{erfc}(r/\sigma)$
and the last term in Eq.~(\ref{Debye_field}). We note that the idea of introducing effective
Debye-H\"{u}ckel, or Yukawa, interactions between polyelectrolyte charges has been widely
used to account for the screening by small ions.\cite{Yukawa1,Yukawa2}
The advantage of our approach, however,
is that it invokes only the long-ranged features of the Debye screening profile, while still
explicitly accounting for the
strong electrostatic core interactions between charges at small distances in the simulations.
This approach should therefore remain accurate for strongly interacting and dense polyelectrolyte systems,
where conventional Debye-H\"{u}ckel  interactions are a very crude approximation.

This work was supported by NSF through grant CHE05-17818 and through TeraGrid resources
provided by the NSCA site under grant CHE070003T.
We are grateful to Jocelyn Rodgers for many helpful remarks.

\bibliography {sources}

\begin{thebibliography}{35}
\expandafter\ifx\csname natexlab\endcsname\relax\def\natexlab#1{#1}\fi
\expandafter\ifx\csname bibnamefont\endcsname\relax
  \def\bibnamefont#1{#1}\fi
\expandafter\ifx\csname bibfnamefont\endcsname\relax
  \def\bibfnamefont#1{#1}\fi
\expandafter\ifx\csname citenamefont\endcsname\relax
  \def\citenamefont#1{#1}\fi
\expandafter\ifx\csname url\endcsname\relax
  \def\url#1{\texttt{#1}}\fi
\expandafter\ifx\csname urlprefix\endcsname\relax\def\urlprefix{URL }\fi
\providecommand{\bibinfo}[2]{#2}
\providecommand{\eprint}[2][]{\url{#2}}

\bibitem[{\citenamefont{Hansen and McDonald}(2006)}]{HansenMcDonald}
\bibinfo{author}{\bibfnamefont{J.-P.} \bibnamefont{Hansen}} \bibnamefont{and}
  \bibinfo{author}{\bibfnamefont{I.~R.} \bibnamefont{McDonald}},
  \emph{\bibinfo{title}{Theory of Simple Liquids}}
  (\bibinfo{publisher}{Academic Press}, \bibinfo{address}{New York},
  \bibinfo{year}{2006}), \bibinfo{edition}{3rd} ed.

\bibitem[{\citenamefont{Hummer et~al.}(1994)\citenamefont{Hummer, Soumpasis,
  and Neumann}}]{ChargedCloud}
\bibinfo{author}{\bibfnamefont{G.}~\bibnamefont{Hummer}},
  \bibinfo{author}{\bibfnamefont{D.}~\bibnamefont{Soumpasis}},
  \bibnamefont{and} \bibinfo{author}{\bibfnamefont{M.}~\bibnamefont{Neumann}},
  \bibinfo{journal}{J. Phys. - Condens. Matt.} \textbf{\bibinfo{volume}{6}},
  \bibinfo{pages}{A141} (\bibinfo{year}{1994}).

\bibitem[{\citenamefont{Wolf et~al.}(1999)\citenamefont{Wolf, Keblinski,
  Phillpot, and Eggebrecht}}]{Wolf}
\bibinfo{author}{\bibfnamefont{D.}~\bibnamefont{Wolf}},
  \bibinfo{author}{\bibfnamefont{P.}~\bibnamefont{Keblinski}},
  \bibinfo{author}{\bibfnamefont{S.~R.} \bibnamefont{Phillpot}},
  \bibnamefont{and}
  \bibinfo{author}{\bibfnamefont{J.}~\bibnamefont{Eggebrecht}},
  \bibinfo{journal}{J. Chem. Phys.} \textbf{\bibinfo{volume}{110}},
  \bibinfo{pages}{8254} (\bibinfo{year}{1999}).

\bibitem[{\citenamefont{Fennell and Gezelter}(2006)}]{Fennell}
\bibinfo{author}{\bibfnamefont{C.~J.} \bibnamefont{Fennell}} \bibnamefont{and}
  \bibinfo{author}{\bibfnamefont{J.~D.} \bibnamefont{Gezelter}},
  \bibinfo{journal}{J. Chem. Phys.} \textbf{\bibinfo{volume}{124}},
  \bibinfo{pages}{234104} (\bibinfo{year}{2006}).

\bibitem[{\citenamefont{Levy and Gallicchio}(1998)}]{ElectEffectsReview}
\bibinfo{author}{\bibfnamefont{R.~M.} \bibnamefont{Levy}} \bibnamefont{and}
  \bibinfo{author}{\bibfnamefont{E.}~\bibnamefont{Gallicchio}},
  \bibinfo{journal}{Annu. Rev. Phys. Chem.} \textbf{\bibinfo{volume}{49}},
  \bibinfo{pages}{531} (\bibinfo{year}{1998}).

\bibitem[{\citenamefont{Bergdorf et~al.}(2003)\citenamefont{Bergdorf, Peter,
  and H\"{u}nenberger}}]{bergdorf:9129}
\bibinfo{author}{\bibfnamefont{M.}~\bibnamefont{Bergdorf}},
  \bibinfo{author}{\bibfnamefont{C.}~\bibnamefont{Peter}}, \bibnamefont{and}
  \bibinfo{author}{\bibfnamefont{P.~H.} \bibnamefont{H\"{u}nenberger}},
  \bibinfo{journal}{J. Chem. Phys.} \textbf{\bibinfo{volume}{119}},
  \bibinfo{pages}{9129} (\bibinfo{year}{2003}).

\bibitem[{\citenamefont{Schreiber and Steinhauser}(1992)}]{SchreiberH}
\bibinfo{author}{\bibfnamefont{H.}~\bibnamefont{Schreiber}} \bibnamefont{and}
  \bibinfo{author}{\bibfnamefont{O.}~\bibnamefont{Steinhauser}},
  \bibinfo{journal}{Biochemistry} \textbf{\bibinfo{volume}{31}},
  \bibinfo{pages}{5856} (\bibinfo{year}{1992}).

\bibitem[{\citenamefont{Chen and Weeks}(2006)}]{LMFPairingAndWalls}
\bibinfo{author}{\bibfnamefont{Y.-G.} \bibnamefont{Chen}} \bibnamefont{and}
  \bibinfo{author}{\bibfnamefont{J.~D.} \bibnamefont{Weeks}},
  \bibinfo{journal}{Proc. Nat. Acad. Sci. USA} \textbf{\bibinfo{volume}{103}},
  \bibinfo{pages}{7560} (\bibinfo{year}{2006}).

\bibitem[{\citenamefont{Rodgers et~al.}(2006)\citenamefont{Rodgers, Kaur, Chen,
  and Weeks}}]{WeeksLMFSim}
\bibinfo{author}{\bibfnamefont{J.~M.} \bibnamefont{Rodgers}},
  \bibinfo{author}{\bibfnamefont{C.}~\bibnamefont{Kaur}},
  \bibinfo{author}{\bibfnamefont{Y.-G.} \bibnamefont{Chen}}, \bibnamefont{and}
  \bibinfo{author}{\bibfnamefont{J.~D.} \bibnamefont{Weeks}},
  \bibinfo{journal}{Phys. Rev. Lett.} \textbf{\bibinfo{volume}{97}},
  \bibinfo{pages}{097801} (\bibinfo{year}{2006}).

\bibitem[{\citenamefont{Ewald}(1921)}]{Ewald}
\bibinfo{author}{\bibfnamefont{P.~P.} \bibnamefont{Ewald}},
  \bibinfo{journal}{Ann. Phys. (Leipzig)} \textbf{\bibinfo{volume}{64}},
  \bibinfo{pages}{253} (\bibinfo{year}{1921}).

\bibitem[{\citenamefont{Darden et~al.}(1993)\citenamefont{Darden, York, and
  Pedersen}}]{MeshEwald}
\bibinfo{author}{\bibfnamefont{T.}~\bibnamefont{Darden}},
  \bibinfo{author}{\bibfnamefont{D.}~\bibnamefont{York}}, \bibnamefont{and}
  \bibinfo{author}{\bibfnamefont{L.}~\bibnamefont{Pedersen}},
  \bibinfo{journal}{J. Chem. Phys.} \textbf{\bibinfo{volume}{98}},
  \bibinfo{pages}{10089} (\bibinfo{year}{1993}).

\bibitem[{\citenamefont{Widmann and Adolf}(1997)}]{Widmann}
\bibinfo{author}{\bibfnamefont{A.~H.} \bibnamefont{Widmann}} \bibnamefont{and}
  \bibinfo{author}{\bibfnamefont{D.~B.} \bibnamefont{Adolf}},
  \bibinfo{journal}{Comput. Phys. Commun.} \textbf{\bibinfo{volume}{107}},
  \bibinfo{pages}{167} (\bibinfo{year}{1997}).

\bibitem[{\citenamefont{Shelley and Patey}(1996)}]{Shelley&Patey}
\bibinfo{author}{\bibfnamefont{J.~C.} \bibnamefont{Shelley}} \bibnamefont{and}
  \bibinfo{author}{\bibfnamefont{G.~N.} \bibnamefont{Patey}},
  \bibinfo{journal}{Mol. Phys.} \textbf{\bibinfo{volume}{88}},
  \bibinfo{pages}{385} (\bibinfo{year}{1996}).

\bibitem[{\citenamefont{Spohr}(1997)}]{Spohr}
\bibinfo{author}{\bibfnamefont{E.}~\bibnamefont{Spohr}}, \bibinfo{journal}{J.
  Chem. Phys.} \textbf{\bibinfo{volume}{107}}, \bibinfo{pages}{6342}
  (\bibinfo{year}{1997}).

\bibitem[{\citenamefont{Yeh and Berkowitz}(1999)}]{EW3DC}
\bibinfo{author}{\bibfnamefont{I.-C.} \bibnamefont{Yeh}} \bibnamefont{and}
  \bibinfo{author}{\bibfnamefont{M.~L.} \bibnamefont{Berkowitz}},
  \bibinfo{journal}{J. Chem. Phys.} \textbf{\bibinfo{volume}{111}},
  \bibinfo{pages}{3155} (\bibinfo{year}{1999}).

\bibitem[{\citenamefont{Arnold et~al.}(2002)\citenamefont{Arnold, de~Joannis,
  and Holm}}]{Holm1}
\bibinfo{author}{\bibfnamefont{A.}~\bibnamefont{Arnold}},
  \bibinfo{author}{\bibfnamefont{J.}~\bibnamefont{de~Joannis}},
  \bibnamefont{and} \bibinfo{author}{\bibfnamefont{C.}~\bibnamefont{Holm}},
  \bibinfo{journal}{J. Chem. Phys.} \textbf{\bibinfo{volume}{117}},
  \bibinfo{pages}{2496} (\bibinfo{year}{2002}).

\bibitem[{\citenamefont{de~Joannis et~al.}(2002)\citenamefont{de~Joannis,
  Arnold, and Holm}}]{Holm2}
\bibinfo{author}{\bibfnamefont{J.}~\bibnamefont{de~Joannis}},
  \bibinfo{author}{\bibfnamefont{A.}~\bibnamefont{Arnold}}, \bibnamefont{and}
  \bibinfo{author}{\bibfnamefont{C.}~\bibnamefont{Holm}}, \bibinfo{journal}{J.
  Chem. Phys.} \textbf{\bibinfo{volume}{117}}, \bibinfo{pages}{2503}
  (\bibinfo{year}{2002}).

\bibitem[{\citenamefont{Weber et~al.}(2000)\citenamefont{Weber, H\"unenberger,
  and McCammon}}]{EwaldArtifacts}
\bibinfo{author}{\bibfnamefont{W.}~\bibnamefont{Weber}},
  \bibinfo{author}{\bibfnamefont{P.}~\bibnamefont{H\"unenberger}},
  \bibnamefont{and} \bibinfo{author}{\bibfnamefont{J.}~\bibnamefont{McCammon}},
  \bibinfo{journal}{J. Phys. Chem. B} \textbf{\bibinfo{volume}{104}},
  \bibinfo{pages}{3668} (\bibinfo{year}{2000}).

\bibitem[{\citenamefont{Weeks}(2002)}]{WeeksLMFReview}
\bibinfo{author}{\bibfnamefont{J.~D.} \bibnamefont{Weeks}},
  \bibinfo{journal}{Annu. Rev. Phys. Chem.} \textbf{\bibinfo{volume}{53}},
  \bibinfo{pages}{533} (\bibinfo{year}{2002}).

\bibitem[{\citenamefont{Weeks et~al.}(1995)\citenamefont{Weeks, Selinger, and
  Broughton}}]{WeeksYBG}
\bibinfo{author}{\bibfnamefont{J.~D.} \bibnamefont{Weeks}},
  \bibinfo{author}{\bibfnamefont{R.~L.~B.} \bibnamefont{Selinger}},
  \bibnamefont{and} \bibinfo{author}{\bibfnamefont{J.~Q.}
  \bibnamefont{Broughton}}, \bibinfo{journal}{Phys. Rev. Lett.}
  \textbf{\bibinfo{volume}{75}}, \bibinfo{pages}{2694} (\bibinfo{year}{1995}).

\bibitem[{\citenamefont{Weeks et~al.}(1998)\citenamefont{Weeks, Katsov, and
  Vollmayr}}]{WeeksYBG2}
\bibinfo{author}{\bibfnamefont{J.~D.} \bibnamefont{Weeks}},
  \bibinfo{author}{\bibfnamefont{K.}~\bibnamefont{Katsov}}, \bibnamefont{and}
  \bibinfo{author}{\bibfnamefont{K.}~\bibnamefont{Vollmayr}},
  \bibinfo{journal}{Phys. Rev. Lett.} \textbf{\bibinfo{volume}{81}},
  \bibinfo{pages}{4400} (\bibinfo{year}{1998}).

\bibitem[{\citenamefont{Katsov and Weeks}(2001)}]{WeeksLMF}
\bibinfo{author}{\bibfnamefont{K.}~\bibnamefont{Katsov}} \bibnamefont{and}
  \bibinfo{author}{\bibfnamefont{J.~D.} \bibnamefont{Weeks}},
  \bibinfo{journal}{J. Phys. Chem. B} \textbf{\bibinfo{volume}{105}},
  \bibinfo{pages}{6738} (\bibinfo{year}{2001}).

\bibitem[{\citenamefont{Chen et~al.}(2004)\citenamefont{Chen, Kaur, and
  Weeks}}]{LMFCoulomb}
\bibinfo{author}{\bibfnamefont{Y.-G.} \bibnamefont{Chen}},
  \bibinfo{author}{\bibfnamefont{C.}~\bibnamefont{Kaur}}, \bibnamefont{and}
  \bibinfo{author}{\bibfnamefont{J.~D.} \bibnamefont{Weeks}},
  \bibinfo{journal}{J. Phys. Chem. B} \textbf{\bibinfo{volume}{108}},
  \bibinfo{pages}{19874} (\bibinfo{year}{2004}).

\bibitem[{\citenamefont{Percus}(1962)}]{Percus}
\bibinfo{author}{\bibfnamefont{J.~K.} \bibnamefont{Percus}},
  \bibinfo{journal}{Phys. Rev. Lett.} \textbf{\bibinfo{volume}{8}},
  \bibinfo{pages}{462} (\bibinfo{year}{1962}).

\bibitem[{\citenamefont{Wu and Li}(2007)}]{DFTreview}
\bibinfo{author}{\bibfnamefont{J.}~\bibnamefont{Wu}} \bibnamefont{and}
  \bibinfo{author}{\bibfnamefont{Z.}~\bibnamefont{Li}}, \bibinfo{journal}{Annu.
  Rev. Phys. Chem.} \textbf{\bibinfo{volume}{58}}, \bibinfo{pages}{85}
  (\bibinfo{year}{2007}).

\bibitem[{\citenamefont{Widom}(1967)}]{Widom}
\bibinfo{author}{\bibfnamefont{B.}~\bibnamefont{Widom}},
  \bibinfo{journal}{Science} \textbf{\bibinfo{volume}{157}},
  \bibinfo{pages}{375} (\bibinfo{year}{1967}).

\bibitem[{\citenamefont{Weeks et~al.}(1971)\citenamefont{Weeks, Chandler, and
  Andersen}}]{WCA}
\bibinfo{author}{\bibfnamefont{J.~D.} \bibnamefont{Weeks}},
  \bibinfo{author}{\bibfnamefont{D.}~\bibnamefont{Chandler}}, \bibnamefont{and}
  \bibinfo{author}{\bibfnamefont{H.~C.} \bibnamefont{Andersen}},
  \bibinfo{journal}{J. Chem. Phys.} \textbf{\bibinfo{volume}{55}},
  \bibinfo{pages}{5422} (\bibinfo{year}{1971}).

\bibitem[{\citenamefont{Kac et~al.}(1963)\citenamefont{Kac, Uhlenbeck, and
  Hemmer}}]{KUH}
\bibinfo{author}{\bibfnamefont{M.}~\bibnamefont{Kac}},
  \bibinfo{author}{\bibfnamefont{G.~E.} \bibnamefont{Uhlenbeck}},
  \bibnamefont{and} \bibinfo{author}{\bibfnamefont{P.~C.}
  \bibnamefont{Hemmer}}, \bibinfo{journal}{J. Math. Phys.}
  \textbf{\bibinfo{volume}{4}}, \bibinfo{pages}{216} (\bibinfo{year}{1963}).

\bibitem[{\citenamefont{Lebowitz et~al.}(1965)\citenamefont{Lebowitz, Stell,
  and Baer}}]{LebStellBaer}
\bibinfo{author}{\bibfnamefont{J.~L.} \bibnamefont{Lebowitz}},
  \bibinfo{author}{\bibfnamefont{G.}~\bibnamefont{Stell}}, \bibnamefont{and}
  \bibinfo{author}{\bibfnamefont{S.}~\bibnamefont{Baer}}, \bibinfo{journal}{J.
  Math. Phys.} \textbf{\bibinfo{volume}{6}}, \bibinfo{pages}{1282}
  (\bibinfo{year}{1965}).

\bibitem[{\citenamefont{Stillinger and Lovett}(1968)}]{SL}
\bibinfo{author}{\bibfnamefont{F.~H.} \bibnamefont{Stillinger}}
  \bibnamefont{and} \bibinfo{author}{\bibfnamefont{R.}~\bibnamefont{Lovett}},
  \bibinfo{journal}{J. Chem. Phys.} \textbf{\bibinfo{volume}{49}},
  \bibinfo{pages}{1991} (\bibinfo{year}{1968}).

\bibitem[{\citenamefont{Ceperley and Chester}(1977)}]{Ceperley}
\bibinfo{author}{\bibfnamefont{D.~M.} \bibnamefont{Ceperley}} \bibnamefont{and}
  \bibinfo{author}{\bibfnamefont{G.~V.} \bibnamefont{Chester}},
  \bibinfo{journal}{Phys. Rev. A} \textbf{\bibinfo{volume}{15}},
  \bibinfo{pages}{755} (\bibinfo{year}{1977}).

\bibitem[{\citenamefont{Rodgers and Weeks}()}]{WaterUnpublished}
\bibinfo{author}{\bibfnamefont{J.~M.} \bibnamefont{Rodgers}} \bibnamefont{and}
  \bibinfo{author}{\bibfnamefont{J.~D.} \bibnamefont{Weeks}}, \bibinfo{note}{in
  preparation}.

\bibitem[{\citenamefont{Denesyuk and Weeks}()}]{PolymerUnpublished}
\bibinfo{author}{\bibfnamefont{N.~A.} \bibnamefont{Denesyuk}} \bibnamefont{and}
  \bibinfo{author}{\bibfnamefont{J.~D.} \bibnamefont{Weeks}}, \bibinfo{note}{in
  preparation}.

\bibitem[{\citenamefont{Micka and Kremer}(1997)}]{Yukawa1}
\bibinfo{author}{\bibfnamefont{U.}~\bibnamefont{Micka}} \bibnamefont{and}
  \bibinfo{author}{\bibfnamefont{K.}~\bibnamefont{Kremer}},
  \bibinfo{journal}{Europhys. Lett.} \textbf{\bibinfo{volume}{38}},
  \bibinfo{pages}{279} (\bibinfo{year}{1997}).

\bibitem[{\citenamefont{Liverpool and Stapper}(1997)}]{Yukawa2}
\bibinfo{author}{\bibfnamefont{T.~B.} \bibnamefont{Liverpool}}
  \bibnamefont{and} \bibinfo{author}{\bibfnamefont{M.}~\bibnamefont{Stapper}},
  \bibinfo{journal}{Europhys. Lett.} \textbf{\bibinfo{volume}{40}},
  \bibinfo{pages}{485} (\bibinfo{year}{1997}).

\end{thebibliography}

\end {document}